\documentclass[10pt]{article}
\usepackage{algorithm}
\usepackage{algpseudocode}
\usepackage{tikz}
\usepackage{forloop}

\newtheorem{lemma}{Lemma}

\newtheorem{theorem}[lemma]{Theorem}
\newtheorem{definition}[lemma]{Definition}

\newcommand{\bigo}{\mathcal{O}}

\newcommand{\fullversion}[1]{}

\usepackage{times}

\usepackage{cite}
\usepackage{graphicx}
\usepackage[cmex10]{amsmath}
\usepackage{array}
\usepackage{mdwmath}
\usepackage{mdwtab}
\usepackage[tight,footnotesize]{subfigure}
\begin{document}
%\title{Finding Associations and Computing Similarity over a Stream of Transactions}

\author{Andrea Campagna and Rasmus Pagh\\
        IT University of Copenhagen, Denmark\\
        Email: {\tt \{acam,pagh\}@itu.dk}}
% \author{\IEEEauthorblockN{Andrea Campagna and Rasmus Pagh}
% \IEEEauthorblockA{IT University of Copenhagen, Denmark\\
% Email: {\tt \{acam,pagh\}@itu.dk}}}


%opening
\title{On Finding Frequent Patterns in Event Sequences%
%\title{Finding Frequent Patterns in RFID Tag Movements%
\thanks{This work was supported in part by the SPOPOS project, supported by the Research and Innovation Agency under the Danish Ministry for Knowledge, Technology and Development.}}

\maketitle

%\IEEEpeerreviewmaketitle

\begin{abstract}
Given a directed acyclic graph with labeled vertices, we consider the problem of finding the most common label sequences (``traces'') among all paths in the graph (of some maximum length $m$). Since the number of paths can be huge, we propose novel algorithms whose time complexity depends only on the size of the graph, and on the frequency $\varepsilon$ of the most frequent traces. In addition, we apply techniques from streaming algorithms to achieve space usage that depends only on $\varepsilon$, and not on the number of distinct traces.

The abstract problem considered models a variety of tasks concerning finding frequent patterns in event sequences. Our motivation comes from working with a data set of 2 million RFID readings from baggage trolleys at Copenhagen Airport. The question of finding frequent passenger movement patterns is mapped to the above problem. We report on experimental findings for this data set.

%Motivated by large databases of RFID tag readings, we consider mining of frequent spatio-temporal patterns. The basic strategy is to convert the problem to discovery of frequent subsequences of readings, with a specified maximum time gap $\Delta$ between subsequent readings. This can lead to a very large number of sequences, and therefore we propose a sampling technique that creates a sample of the set $S_m$ of all sequences without generating it explicitly. By choosing a suitable sampling probability, frequent patterns from $S_m$ are also frequent in the sample with high probability. Finally, we apply data streaming algorithms to compute the frequent patterns in small space.

%We perform experiments with the proposed method on a data set of 2 million RFID readings from baggage trolleys at Copenhagen Airport, and present some of our findings.
\end{abstract}


%%%%%%%%%%%%%%%%%%%%%%%%%%%%%%%%%%%%%%%%%%%%%%%%%%%%%%%%%%%%%%%%%%%%%%%%%%%%%%%%%

\section{Introduction}

Sequential pattern mining has attracted a lot of interest in recent years. However, some of the probabilistic techniques that have proven their efficiency in mining of frequent itemsets have, to our best knowledge, not been transferred to the realm of sequence mining. The aim of this paper is to take a step in that direction, namely, we propose an analogue of Toivonen's sampling-based algorithm for frequent itemset mining~\cite{toivonen} in the context of sequential patterns.

At a conceptual level we work with a new, simple formulation of the problem: The input is a directed acyclic graph (DAG) where the vertices are events and there is an edge between two events if they are considered to be connected (i.e., part of the same event sequences). Vertices are labeled by the type of event they represent. This allows certain flexibility in modeling that is lacking in many other formulations:
\begin{itemize}
\item Spatio-temporal events can be connected based on both spatial and temporal closeness.
\item Events that have an associated time range (rather than a single time stamp) can be connected based on an arbitrary closeness criterion.
\end{itemize}

The data mining task we consider is to find the most common sequences of event types (``traces'') among all paths in the DAG, or more generally all paths of some maximum length $m$. The challenge is to handle the huge number of paths that may be present in a DAG. 

\paragraph{Example} Consider data on the history of URLs visited by a user, where each URL is labeled by its domain name. If she visits the domains {\tt www.techcrunch.com}, {\tt www.oracle.com}, and {\tt www.itu.dk} in this order, there may be a connection between the first and second site, and between the second and third site. If all visits happen within a few minutes one could also imagine that the second site was merely a detour, and there is a connection from the first to the third site. This is naturally modeled using a graph having URL visits as vertices, and directed edges between vertices that we deem connected (based on any criterion, e.g., temporal closeness). We label vertices by domain name, and look for frequently occurring label sequences, {\em traces\/}, on paths in the graph.
\vspace{-5mm}

\begin{center}
\usetikzlibrary{positioning} 
\begin{tikzpicture}[scale=.7] 
\tikzstyle{v}=[circle,minimum size=1mm,draw,thick];
\node[v] (z) {{\tt{\small goo}}}; 
\node[v] (a) [right=of z] {{\tt{\small tec}}}; 
\node[v] (b) [right=of a] {{\tt{\small ora}}}; 
\node[v] (c) [right=of b] {{\tt{\small itu}}}; 
\node[v] (d) [right=of c] {{\tt{\small goo}}}; 
\draw[thick,->] 
(z) to node {} (a);
\draw[thick,->,out=-25,in=-155] 
(z) to node {} (b);
\draw[thick,->,out=25,in=155] 
(a) to node {} (c);
\draw[thick,->] 
(a) to node {} (b);
\draw[thick,->] 
(b) to node {} (c);
\draw[thick,->] 
(c) to node {} (d);
\end{tikzpicture}
\end{center}

We might be interested in such frequent event sequences for a variety of reasons, e.g.~improved understanding of browsing behavior for advertisers (avoid paying for many page impressions to the same user), and page recommendations (``users who visited the same sequence of domains as you, often went on to the domain\dots''). We should be able to detect the connection between sites even if they are not visited in succession. For example, many browsing histories will interleave visits to hubs such as {\tt google.com} and {\tt yahoo.com} with visits to topic specialized domains.

\subsection{Approach}

We start from the observation that the number of paths in a DAG can be extremely large, even if the path length is restricted to some small number $m$. 
For example, the DAG pictured below has 16 vertices and 45 edges, but the number of paths is 10919.

% Python function:
% def path(N):
%...   if(N==1): return 1;
%...   if(N==2): return 2;
%...   if(N==3): return 3;
%...   return 1+path(N-1)+path(N-2)+path(N-3);
%... 
%>>> print path(16);
%10935
%Subtracted all 16 paths with only one node.

\begin{center}
\begin{tikzpicture}[scale=0.55] 
\tikzstyle{v}=[circle, minimum size=2mm,inner sep=0pt,draw] 
\foreach \i in {1,...,16} {
\node[v] 
(G-\i) at (\i,0) {};
} 
\foreach \i/\j in {1/2,2/3,3/4,4/5,5/6,6/7,7/8,8/9,9/10,10/11,11/12,12/13,13/14,14/15,15/16} {
\draw[->,out=0,in=180] (G-\i) to node {} (G-\j);
}
\foreach \i/\j in {1/3,2/4,3/5,4/6,5/7,6/8,7/9,8/10,9/11,10/12,11/13,12/14,13/15,14/16} {
\draw[->,out=30,in=150] (G-\i) to node {} (G-\j);
}
\foreach \i/\j in {1/4,2/5,3/6,4/7,5/8,6/9,7/10,8/11,9/12,10/13,11/14,12/15,13/16} {
\draw[->,out=-30,in=-150] (G-\i) to node {} (G-\j);
}

\end{tikzpicture}
\end{center}

More generally, we expect the number of paths to increase exponentially with $m$.
In our experiments we see that, even for small $m$, the number of paths is much larger than the size of the DAG.

Our algorithm rests on a novel {\em sampling\/} procedure that is able to create a sample of any desired size, in time that is linear in the size of the DAG (for preprocessing) and the size of the sample. 
This allows a time complexity for the mining procedure that depends only on the {\em frequency\/} $\varepsilon$ of the most common traces, rather than the total number of traces. 
We also apply a technique from data streaming algorithms to achieve space that depends on $\varepsilon$ rather than on the number of distinct traces.

Though our formulation does not capture all the many aspects present in other approaches to sequential pattern mining, we believe that it possesses an attractive combination of {\em expressive modeling\/} and {\em algorithmic tractability}. 
%In addition, our algorithms can likely be extended to cover 

%%%%%%%%%%%%%%%%%%%%%%%%%%%%%%%%%%%%%%%%%%%%%%%%%%%%%%%%%%%%%%%%%%%%%%%%%%%%%%%%%

\subsection{Problem definition}

We are given a directed acyclic graph $G=(V,E)$, and a function label$(v)$ that returns the label of a vertex. A path $p$ in $G$ is a sequence of vertices $v_1,v_2,\dots,v_j\in V$ such that $(v_i,v_{i+1})\in E$ for $i=1,\dots,j-1$.
A path $p$ has a {\em trace} $\text{label}(p)$, which is the vector of labels on the path. Let $S_m$ be the multiset of all path traces of length at most $m$, i.e.,
$$S_m = \{ \text{label}(p) \; | \; p \text{ is a path in $G$ of length at most $m$}\} \enspace .$$
The data mining task is to find the most frequent traces in $S_m$. It comes in several flavors:
\begin{itemize}
\item {\bf Top-$k$}. For a parameter $k$, find the $k$ traces that have the most occurrences in~$S_m$ (breaking ties arbitrarily).
\item {\bf Frequency $\varepsilon$}. Find the set of traces that have frequency $\varepsilon$ or more in $S_m$.
\item {\bf Monte Carlo}. For both the above variants we can allow an error probability $\delta$ (typically allowing a false negative probability, i.e., that we fail to report a trace with probability $\delta$).
\end{itemize}
In this paper emphasis will be on Monte Carlo algorithms for the frequency variant. However, we note that one can also obtain results for top-$k$ by a simple reduction.

\subsection{Related work}

There is a large body of related work on sequential pattern mining, see e.g.~\cite{journals/datamine/MannilaTV97,DBLP:conf/edbt/SrikantA96,joshi,HSDJTT,ZB_tr_2003,pisa,Chen20061203,PSpan}. These works deviate from the present one in that they consider the input as a sequence of timestamped events, and allow a host of formulations of what kinds of subsequences are of interest. 
In contrast, we put the modeling of interesting subsequences into the description of the event sequence (by defining DAG edges), and the patterns sought are simple strings. 
This allows us to do things that we believe have not been done, and are probably difficult, in traditional sequential data mining settings, namely making use of sampling methods.

The difficulty with sampling is that patterns can overlap in many ways, so any straightforward approach will fail to produce a sample that correctly ``represents'' the original data.
As an example, suppose that the pattern $\text{\tt a}^{2m}$ occurs in the input, which means $k+1$ occurrences of $\text{\tt a}^{m}$. 
If we sample events with probability 50\%, the probability that an occurrence of $\text{\tt a}^{m}$ remains in the sample is 1/2. On the other hand, if there are $k+1$ non-overlapping occurrences of $\text{\tt a}^{m}$, the probability that this is seen in the sample may be much lower. For example, for the string $(\text{\tt a}^{m}\text{\tt b}^{m})^{m+1}$ the probability is $O(m/2^m)$, i.e., exponentially decreasing as~$m$ grows. This means that there is no direct way of going from the number of occurrences in the sample to the number of occurrences in the original string.

Similar problems make use of sampling methods in general graph mining difficult. Suppose that we sample vertices (or edges) with probability $p$. If all triangles in a graph overlap in a single vertex, the sample will contain no triangles at all with probability $1-p$. On the other hand, if there is the same number of vertex (edge) disjoint triangles, we are likely to sample close to a fraction $p^3$ of them. As before, we cannot estimate the number of occurrences in the original graph based on the number of occurrences in the sample.

% For example, suppose that our graph contains a clique of size $2k$, then there are at least $\binom{2k}{k}$ (overlapping) cliques of size $k$. If this number is above the support threshold, such that the $k$-clique is to be reported,

%All characterized by windowing and/or time gap constraints. All do exact counting of patterns/episodes. Lacking analogue of Toivonen's algorithm.

% {\bf XXX Andrea, please help. Need to include citations of (at least) these at XXX above:
% 
% Srikant Agrawal, Mining sequential patterns: Generalization...
% 
% Joshi, M. V., Karypis, G., and Kumar, V. (2001). A universal formulation of sequential patterns. In Proceedings of the KDD’2001 workshop on Temporal Data Mining, San Fransisco, August 2001.
% CDIST\_O counting scheme
% 
% Discovering Sequential Association Rules with Constraints and Time Lags in Multiple Sequences
% 
% Sequential Pattern Mining: A Survey
% 
% Mining sequences with temporal annotations
% 
% Constraint-based sequential pattern mining: The consideration of recency and …
% 
% PrefixSpan. PrefixSpan [Pei et al. 2001]}
% DONE

%"Sequential pattern mining is used in a great spectrum of areas. In computational biology, sequential pattern mining is used to analyze the mutation patterns of dif- ferent amino acids. Business organizations use sequential pattern mining to study customer behaviors. Sequential pattern mining is also used in system performance analysis and telecommunication network analysis."

%%%%%%%%%%%%%%%%%%%%%%%%%%%%%%%%%%%%%%%%%%%%%%%%%%%%%%%%%%%%%%%%%%%%%%%%%%%%%%%%%

\section{Our solution}

\subsection{Generation of all traces}

As a warmup we consider the task of producing the multiset $S_m$ of all traces having maximum length $m$.
We will use the notation $S_i(v)$ to denote the multiset of traces corresponding to  paths (of length at most $i$) starting in node $v$. Clearly $S_0(v)=\emptyset$. For $i>0$ we have the recursive definition
$$S_i(v)=\{\text{label}(v)\} \times (\epsilon \cup \bigcup_{v',\, (v,v')\in E} S_{i-1}(v')),$$
where $\epsilon$ denotes the empty trace (note that this symbol is different from $\varepsilon$ denoting the frequency), and $\bigcup$ is multiset union. Clearly we have $S_m = \bigcup_{v\in V} S_m(v)$.

These equalities lead to a simple recursive algorithm, shown in Figure~\ref{fig:naive}. It is easy to see that if traces are represented in a reasonable way (e.g.~as singly linked lists) the running time is linear in the size $|V|+|E|$ of the graph and the total length of the traces generated.

{\bf Succinct output.} If we are satisfied with returning hash values of the traces (unique with high probability) the time can be improved such that only $\bigo(1)$ time is used for each trace, i.e.~time $\bigo(|V|+|E|+|S_m|)$ in total. This can be done using a standard incremental string hashing method such as Karp-Rabin~\cite{KarpRabin}. Observe that the output is sufficient to find the {\em hash values\/} of the most frequent traces in $S_m$ (with a negligible error probability). A second run of the procedure could then output the actual frequent traces, e.g.~by looking up the count of each hash value computed.

%\medskip

\begin{figure}
\begin{algorithmic}[1]
\Procedure{AllTraces}{$v,t,i$}
\If{$i>0$}
\State{{\bf output} $t || \text{label}(v)$}
\For{{\bf each} $v'$ where $(v,v')\in E$}
\State{{\sc AllTraces}$(v', t || \text{label}(v),i-1)$}
\EndFor
\EndIf
\EndProcedure
\medskip
\For{$v\in V$} 
\State{{\sc AllTraces}$(v,\epsilon,m)$}
\EndFor
\end{algorithmic}
\caption{The procedure {\sc AllTraces} outputs the concatenation of a trace prefix $t$, and each trace starting at $v$ having length at most $i$. The notation $||$ is for concatenation of traces. Lines 7--9 call {\sc AllTraces} for all vertices $v$, with the empty trace $\epsilon$ as prefix, producing the multiset $S_m$ of all traces of length at most $m$.}\label{fig:naive}
\end{figure}

%\medskip

\subsection{Generation of a random sample}

If the patterns we are interested in occur many times, substantial savings in time can be obtained by employing a sampling procedure. That is, rather than generating $S_m$ explicitly we are interested in an algorithm that produces each trace in $S_m$ with a given probability $p$, independently. This will reduce the expected number of samples to a fraction $p$ of the original. The choice of $p$ is constrained by the fact that we still want to sample each frequent trace a fair number of times (to minimize the probability of {\em false negatives\/} being introduced by the sampling).

\paragraph{Counting phase}
Our algorithm starts by computing, for $i=1,\dots,m$ the number of paths $v.c[i]$ of length at most $i$ that start in each vertex~$v$. We assume that this can be done using standard precision (e.g.~64 bit) integers. The algorithm shown in Figure~\ref{fig:counttrace} mimics the structure of the na\"ive generation algorithm, but uses memoization (aka.~dynamic programming) to reduce the running time.

For each $i\leq m$ the cost of all calls to {\sc CountTraces} with parameters $(v,i)$, disregarding the cost of recursive calls, is easily seen to be proportional to the number of edges incident to $v$. This means that the total time complexity of the counting phase is $\bigo(|E| m)$. The space usage is dominated by an array of size $m$ for each vertex, i.e., it is $\bigo(|V| m)$.

\begin{figure}
\begin{algorithmic}[1]
\Function{CountTraces}{$v,i$}
\If{$v.c[i] = ${\bf null}}
\State{$v.c[i]\leftarrow 1$}
\For{{\bf each} $v'$ where $(v,v')\in E$}
\State{$v.c[i]\leftarrow v.c[i] +${\sc CountTraces}$(v',i-1)$}
\EndFor
\EndIf{}
\State{\Return $v.c[i]$}
\EndFunction
\medskip
\For{$v\in V$} 
\State{{\sc CountTraces}$(v)$}
\EndFor
\end{algorithmic}
\caption{Recursive computation of the paths of traces for each starting vertex, using memoization. The algorithm assumes that each value $v.c[0]$ is initially set to zero, and each value $v.c[i]$, $0<i\leq m$, is initially {\bf null}.}\label{fig:counttrace}
\end{figure}

\paragraph{Sampling phase}

Consider the multiset $S_i(v)$ of traces, which has size $v.c[i]$ by definition. The probability that none of these traces are sampled should be $(1-p)^{v.c[i]}$. Conditioned on the event that at least one trace from $S_i(v)$ is sampled, we either have to sample a trace of length more than one (starting with label$(v)$), or include the trace $\{v\}$ in the sample. In a nutshell, this is what the procedure {\sc SampleTraces} of Figure~\ref{fig:sampletrace} does.

Let rand() denote a function the returns a uniformly random number in $[0;1]$, independently of previously returned values. The condition $\text{rand}()>(1-p)^{v.c[m]}$ holds with probability $1-(1-p)^{v.c[m]}$, so lines 14--16 call {\sc SampleTraces}
if and only if we need to sample at least one trace from $S_m(v)$. In the procedure {\sc SampleTraces} we use, similarly to above, a parameter $t$ to pass along a trace prefix. The variable $out$ is used to keep track of whether a trace has been output in the recursive calls. If $out$ is false after all recursive calls we sample $t || \text{label}(v)$. For each $v'$ with $(v,v')\in E$ the probability that we do {\em not\/} sample any trace from $\text{label}(v) || S_{i-1}(v')$ is $(1-p)^{v'.c[i-1]}/(1-(1-p)^{v.c[i]})$. This is exactly the correct probability since we condition on at least one trace in $S_i(v)$ being sampled.
% XXX independence?

\begin{figure}
\begin{algorithmic}[1]
\Procedure{SampleTraces}{$v,t,i$}
\State{$out \leftarrow false$}
\For{{\bf each} $v'$ where $(v,v')\in E$}
\If{rand()$>(1-p)^{v'.c[i-1]}/(1-(1-p)^{v.c[i]})$}
\State{{\sc SampleTraces}$(v', t || \text{label}(v), i-1)$}
\State{$out \leftarrow true$}
\EndIf
\EndFor
\If{$out=false$ {\bf or} rand()$<p$}
\State{{\bf output} $t || \text{label}(v)$}
\EndIf
\EndProcedure
\medskip
\For{$v\in V$}
\If{rand()$>(1-p)^{v.c[m]}$}
\State{{\sc SampleTraces}$(v,\epsilon,m)$}
\EndIf
\EndFor
\end{algorithmic}
\caption{The procedure {\sc SampleTraces} outputs the concatenation of a trace prefix $t$ and a random sample of the traces starting at $v$ of length at most $i$. The traces are sampled from the conditional distribution that is guaranteed to sample at least one trace.
As before, the notation $||$ is for concatenation of traces, and $\epsilon$ denotes the empty trace. Lines 13--17 call {\sc SampleTraces} for each vertex $v$ with probability $1-(1-p)^{v.c[i]}$, to produce a sample of all traces starting at $v$ having length at most $i$, where each trace is chosen independently at random with probability~$p$.}\label{fig:sampletrace}
\end{figure}

{\bf Refinement.} Observe that the probability in line 4 may be precomputed for each edge and value of $i$. Even with this optimization, a direct implementation of the pseudocode in Figure~\ref{fig:sampletrace} may spend a lot of time in the {\bf for} loop of {\sc SampleTraces} without producing any output. To get a theoretically satisfying solution we may preprocess, for each $(v,i)$, the probabilities $p_1,p_2,\dots,p_d$ of making the recursive calls. Specifically, for $j=0,\dots,d$ we consider the probabilities $q_j=\Pi_{j'\leq j} (1-p_{j'})$ that no recursive call is made in the first $j$ iterations. If we choose $r$ uniformly at random in $[0;1]$ then the probability that $q_{j-1} > r > q_{j}$ is exactly the probability that the first recursive call is in the $j$th iteration. Similarly, the probability that $r>q_d$ is exactly the probability that no recursive call is made. Thus, by doing a binary search for $r$ over $q_d,\dots,q_0$ we may choose, with the correct probability, the first iteration $j_1$ in which there should be a recursive call. The same method can be repeated, using a random value $r$ in $[0;q_{j_1}]$ to find the next recursive call, and so on.

In the worst case this uses time $\bigo(\log |V|)$ per recursive call. We can exploit the fact that we are searching for a random value $r$ to decrease this to $\bigo(1)$ expected time. The idea is to represent the values $q_j$ in a binary trie that is precomputed for each node. In addition we store for each string $s\in \{0,1\}^{\lceil\log d\rceil}$ a pointer to the node in the trie that corresponds to the longest prefix of $s$. The number of bits of $r$ needed to determine its position in $q_d,\dots,q_0$ is at most $\lceil \log d\rceil + t$ with probability at least $1-2^{-t}$. Using the pointers we can thus in expected time $\bigo(1)$ find the node in the trie that has the longest common prefix with the binary representation of $r$. This, in turn, determines the rank of $r$ in $q_d,\dots,q_0$.

As before, we can choose to have a succinct output where traces are represented by the hash values of their traces, with no increase in time complexity.

\subsection{Time and error analysis}

For the time analysis we focus on the refined implementation described above, since it allows a clean and exact theoretical analysis. A similar analysis of the version stated in the pseudocode can be made under the assumption that the outdegree of vertices in $G$ is bounded by a constant. Observe that if {\sc SampleTraces} makes $c$ recursive calls this takes expected time $O(1+c)$. Also observe that the total number of procedure calls is upper bounded by the total length of all sampled traces --- this is because each recursive call is guaranteed to produce at least one output. Combining these facts we see that the expected time for all calls to {\sc SampleTraces} is linear in the length $\ell$ of all traces sampled. Notice that the expected value of $\ell$ is $\bigo(p |S_m| m)$. Since $\ell$ is independent of the random choices determining the running time of the data structure in the refined implementation we can conclude that the total expected running time of the code in Figures~\ref{fig:counttrace} and~\ref{fig:sampletrace} is $\bigo(|V| + |E| m + p |S_m| m)$.

The parameter $p$ must be chosen such that $p = C/\varepsilon$, where $C>1$ is a parameter that determines the false negative probability. The expected number of times that we sample a trace with frequency $\varepsilon'$ is $C\varepsilon'/\varepsilon$, and since the samples are independent, the number of samples follows a binomial distribution. By Chernoff bounds, this means that if $\varepsilon'\geq \varepsilon$ then the number of samples is at least $C/2$ with probability $1-2^{-\Omega(C)}$. Examples of concrete error probabilities are given in our experimental section. We have the following theoretical result:

\begin{theorem}
We can generate a random sample of $S_m$ in expected time $\bigo(|V|+|E|m+\log(1/\delta)/\varepsilon)$ such that any trace with frequency $\varepsilon$ or more has frequency at least $\varepsilon / 2$ in the random sample with probability $1-\delta$.\hfill $\circ$
\end{theorem}

\medskip

Observe that the running time is independent of the total number of traces in $S_m$.

\subsection{Putting things together}

It remains to assess how to choose, among the samples, the ones that are actually interesting. In particular, we are interested in those traces appearing in the
sample at least $C / 2$ times.

This problem can be efficiently faced using a \textit{frequent items}
algorithm. Such algorithms are widely used in data streaming
contexts, and guarantee very small space usage. 
%and very high probability
%of reporting only the interesting (i.e., frequent) elements.
A comprehensive treatment and an experimental comparison between various techniques can be found in~\cite{journals/pvldb/CormodeH08}.
%The problem itself dates back at least to the 1980s, and can be formalized
%in this way:
\begin{definition}\label{prob:freqPairs}
 Given a stream $\mathcal{S}$ of $n$ elements, a frequency threshold
 $\eta$, and let $f_i$ be the the frequency of $i$ in $\mathcal{S}$.
The {\em frequent items\/} problem consists in returning a set $\mathcal{F}$ of size at most $1/\eta$ such that for all $i$ with $f_i > \eta$, $i\in \mathcal{F}$.\hfill $\circ$
\end{definition}

\medskip

Observe that false positives, with $f_i < \eta$, can appear in the output.
To eliminate these, we simply make another pass (i.e., generate the same sample again) to compute exact frequencies.
\begin{theorem}
 Given a stream of elements representing the set of samples of traces
 produced by \textsc{SampleTraces}, the space needed in order to output
 the traces with frequency at least $\varepsilon / 2$, without
 producing any trace with frequency less than $\varepsilon / 2$, is
 $\bigo(1/\varepsilon)$ words.\hfill $\circ$
\end{theorem}

\section{From event sequence to a DAG}

An event sequence is a set $S$ of tuples of the form $(t,i,\ell)$, where $t\in \mathbf{R}$ is a time stamp, $i$ is a tag identifier, and $\ell$ is a label (in our application case of RFID readings from baggage trolleys, $i$ identifies the RFID on a trolley and $\ell$ is a location identifier that indicates an approximate location, namely vicinity of an antenna,  of $i$ at time $t$). In this work we do not consider the physical locations of antenna as part of the input. %Instead, we assume that each time a tag is read this may generate multiple tuples $(t,i,\ell_1), (t,i,\ell_2),\dots$ where $\ell_1,\ell_2,\dots$ are location identifiers at different granularities. For example, in an airport $\ell_1$ may denote the ID of the antenna that did the reading, $\ell_2$ may denote the local area of the antenna (e.g.~``central shopping street''), and $\ell_3$ may denote the terminal of the antenna. The definition of this hierarchy of locations is part of the input: For each antenna the user must specify the list of location identifiers that a reading should generate.

Formally we may define the problem as follows: For a given number $\Delta$, the input set specifies a directed acyclic graph $G_\Delta = (V,E_\Delta)$, where each observation is a vertex, and there is an edge from $v_1$ to $v_2$ if and only if the vertices are observations of the same tag, at different locations, separated by at most $\Delta$ time units (we use minutes as the time unit from now on). 

To produce the DAG we sort the data by tag ID and timestamp. Note that this makes it easy to find all the edges from a particular vertex $v$ in $G_\Delta$: Simply scan the sorted list forward until either the timestamp differs by more than $\Delta$ from that of $v$, or we reach a node corresponding to another tag. 

{\bf Example.} If $\Delta=20$ and we observe locations 1, 2, 3, 6, 7 at time 10, 20, 30, 60, 70, the following subsequences are considered to reflect a movement: 1-2, 2-3, 1-2-3, 1-3, 6-7. Notice the inclusion of 1-3, where one observation is skipped, since there is at most $\Delta$ minutes between the observation of 1 and 3. \hfill $\circ$

\section{Experiments}

We have worked with a data set consisting of readings of RFID (Radio-Frequency ID) tags by fixed-position antenna. RFID chips can be identified only when they are in the proximity of an antenna, which means that readings give approximate information about the location of an RFID tag. Such data sets, as well as similar data sets based on other technologies, are becoming increasingly available as more and more items, from parcels to items in shops, are being tagged with RFID chips.

In order to construct the DAG, we have cleaned some of the noise present
in the data.
One source of noise was due to the presence of sequences of readings
regarding trolleys remaining in zones where the range of two antennas
is overlapping.
This sequences of alternating readings had the form $(x^+y^+)(x^+y^+)^+$.
In order to clean up this interferences, we replaced the elements of such
a kind of sequences, using a new zone label that represents the
zone of overlap of the range of antennas. In particular we have used,
for a sequence $(x^+y^+)(x^+y^+)^+$, the label $\min\{x,y\}*100 + \max\{x,y\}$. 

Notice that this can be thought as an increase in the resolution
of the readings, making the granularity of the information finer.
In some sense this modification allows for a cleaner sight on the
movement of some trolleys.

Another source of noise, sometimes connected with the one just described,
is the presence of sequences of readings regardings the same zone for a
given trolley. In order to avoid having traces of the form $t=(Vyy^+W)$,
where $V$ and $W$ are sequences of readings, we considered only one
occurrence of $y$, properly managing the timestamps of the readings.
In particular this means that, assuming the difference
in time between any two consecutive $y$ is within the threshold $\Delta$,
in the DAG we put a directed edge $(v,y)$, $v \in V$ iff the first
occurrence of $y$ after $V$ occurred within time $\Delta$ from
$v$. Moreover we put a directed edge $(y,w)$, $w \in W$ iff $w$ happened
within time $\Delta$ from the last reading of $y$ in $t$.

% \paragraph{Motivation}
% We believe that the abstract problem we consider may be of interest in many scenarios where data is a sequence of events (or more generally, a set of events with no cyclic connections). However, our concrete motivation and experimental results come from work with spatio-temporal data arising from passenger movement patterns in an airport.
% 
% We have worked with a data set consisting of readings of RFID (Radio-Frequency ID) tags by fixed-position antenna. RFID chips can be identified only when they are in the proximity of an antenna, which means that readings give approximate information about the location of an RFID tag. Such data sets, as well as similar data sets based on other technologies, are becoming increasingly available as more and more items, from parcels to items in shops, are being tagged with RFID chips. In addition to tracking of individual items (which is often the primary reason for investing in RFID technology), this opens up the possibility of exploring the movement patterns recorded. Since such data sets may be large, and due to the inherent errors and uncertainties in position data obtained using RFID, it makes sense to consider aggregated features rather than individual tag trajectories. 
% %Instead of considering spatio-temporal queries on such data, we consider {\em data mining\/} techniques that find the ``most frequent'' patterns in the data set. 

\begin{figure}
\begin{center}
\includegraphics[width=0.8\linewidth]{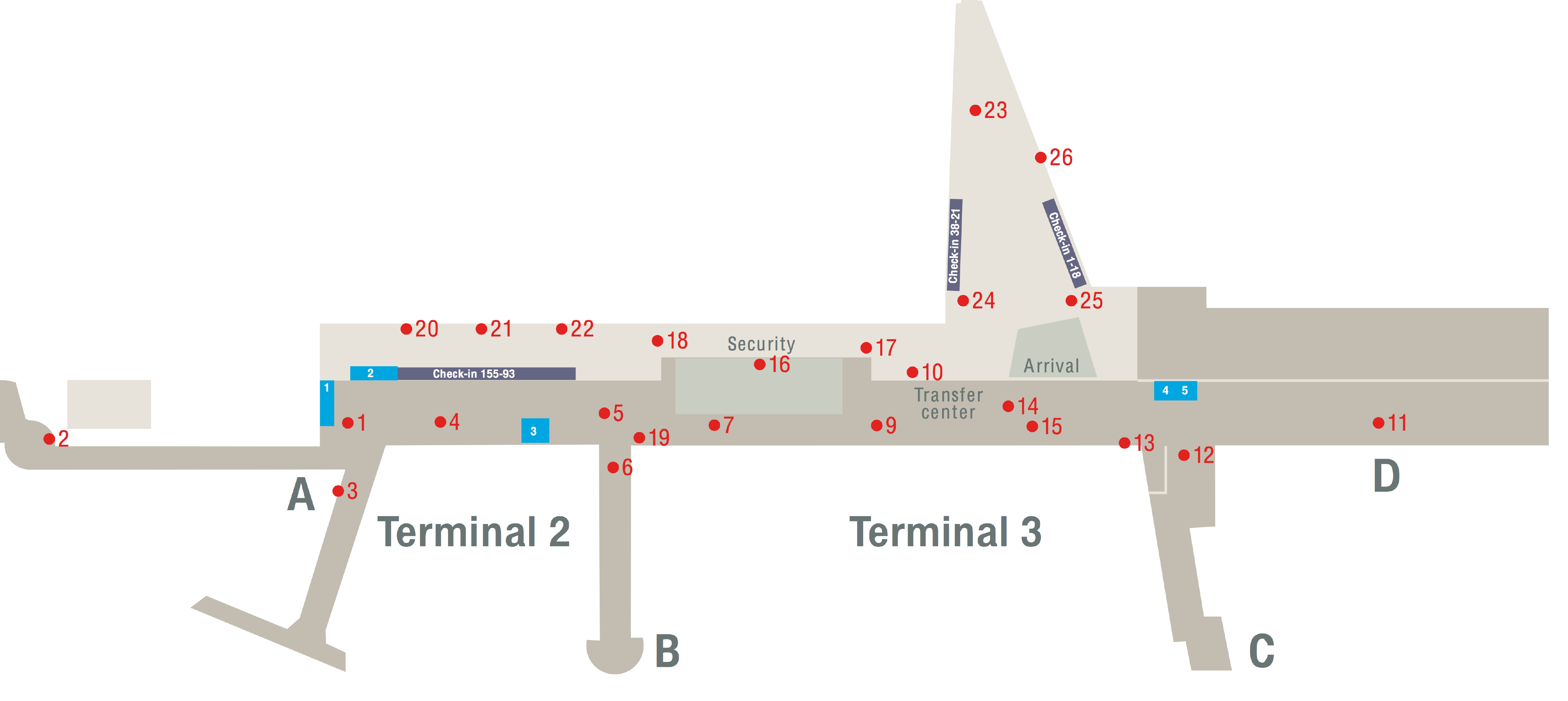}
\end{center}
\caption{RFID antenna in Copenhagen Airport. }
\end{figure}

% \subsection{Problem definition}
% 
% Let $\Delta$ be an integer parameter. For each tag identifier $i$ we use a heuristic to connect observations: If $i$ is observed at different locations within a time span of $\Delta$ minutes we assume that this is due to a passenger moving from one location to another. On the other hand, if two consecutive readings for some tag identifier differ by more than $\Delta$ minutes we assume that the readings are not part of the same movement.
% 
% {\bf Example.} If $\Delta=20$ and we observe locations 1, 2, 3, 6, 7 at time 10, 20, 30, 60, 70, the following subsequences are considered to reflect a movement: 1-2, 2-3, 1-2-3, 1-3, 6-7. Notice the inclusion of 1-3, where one observation is skipped, since there is at most $\Delta$ minutes between the observation of 1 and 3.

%%%%%%%%%%%%%%%%%%%%%%%%%%%%%%%%%%%%%%%%%%%%%%%%%%%%%%%%%%%%%%%%%%%%%%%%%%%%%%%%%

% Number of edges:

% java P_Subsets_generator ../../bin/newDump 20 2 | wc -l
% 4059250
%
%$ java P_Subsets_generator newDump 10 2 | wc -l
%2657931
%
%$ java P_Subsets_generator newDump 5 2 | wc -l
%1721448
%
%$ java P_Subsets_generator newDump 3 2 | wc -l
%1228759

% Number of vertices:
% wc -l SPOPOS/bin/newDump 
% 2206302 SPOPOS/bin/newDump

%For the experiments we have used the dataset described before. The
%reason why we used such a dataset has to be found in the fact that it
%suited quite well the needs of the abstract formulation of the problem,
%and was massive enough to be challenging for our algorithm.

%Moreover, we had not any direct DAG data.
It is necessary to point out that our method differs from the previous
approaches in the way we look for frequent patterns. This means that our results are not
directly comparable with the ones that can be found in literature, so we do not compare to existing algorithms.

\subsection{Results}

We ran a set of experiments on the airport data, in order to understand how many
patterns would have been generated for a given $\Delta$ and a size $m$. Figure~\ref{fig:results} shows the size of the graph for different sizes of $\Delta$.
\begin{figure}
 \begin{center}
  \begin{tabular}{|ccc|}
   \hline
    $\Delta$ & |V| & |E| \\
   \hline
      20 & 2206302 & 4059250\\
      10 & 2206302 & 2657931\\
       5 & 2206302 & 1721448\\
       3 & 2206302 & 1228759\\
  \hline
  \end{tabular}
\caption{Size of the airport DAG for different values of~$\Delta$. As can be seen all graphs are quite sparse, and in fact many nodes have no outgoing edges. This is due to a relatively low resolution in the data set.}
\label{fig:dagsize}
 \end{center}
\end{figure}
We compare the obtained results with the expected performance of our
algorithm.

Figure~\ref{fig:results} reports some interesting characteristics of
the data when fixing $\Delta$ and $m$. In particular the table contains
the number of traces generated, the frequency of the $100$th most frequent 
trace and the ratio between the space needed in case of an
exact computation and the space required when our algorithm is used. Note that the space to represent the DAG and the counts is not counted in this ratio. The rationale for this is that as we consider longer event sequences the space for the DAG representation is expected to become negligible compared to the space needed for finding the most common traces. 
% XXX We really should try to figure out if this is true...
%
\begin{figure}
 \begin{center}
  \begin{tabular}{|cccccc|}
   \hline
    $\Delta$ & $m$ & Tot. traces & Dis. traces &top $100$th & ratio\\
   \hline
      20 & 5 & 365818472 & 4311942 & 168000 & 990\\
      10 & 5 & 106678064 & 1712646 & 52951 &  425\\
      10 & 3 & 6196850 & 50085 & 9458 & 38.2\\
       5 & 5 & 66947355 & 631300 & 42008 & 198\\
       3 & 5 & 23152990 & 280454 & 15363 & 93\\
  \hline
  \end{tabular}
\caption{Characteristics of the data for several
 combinations
 of $\Delta$ and~$m$. The third column, Tot.\@ traces, represents the total
 number of traces that would be generated by the na\"\i ve approach; the
 Dis.\@ traces column represents the number of distinc traces; the top
$100$th column contains the frequency of the $100$th most frequent trace;
the column ratio represents the saving we would achive using a frequency threshold equal to the one represented in the top $100$th column.}\label{fig:results}
 \end{center}
\end{figure}

From the results of the test it is clear that great savings can be
achieved when the frequencies we are interested in are not too low.
In a case, nearly 3 orders of magnitude of space can be saved using our
approach.
As a matter of fact, when we are interested in very frequent traces, and
this is often the case in many practical applications, the sampling outputs
a large number of samples for each interesting trace, so
that a low sampling ratio can be used.

Figure~\ref{fig:time} shows the number of samples we would take in 
expectation when $C=10$ is used. The table gives the flavor of the saving
in time that could be achieved with respect to generating all the possible
traces. Here we notice that the total number of traces is already 1--2 orders of magnitude larger than the size of the DAG, so we expect an improvement in running time of at least 1 order of magnitude. Larger values of $C$ will increase the running time proportionally, but decrease the error probabilities. Table~\ref{table:errorprob} shows false negative probabilities, as well as probabilities that traces with frequency below $\varepsilon /4$ are reported.

\begin{figure}
 \begin{center}
  \begin{tabular}{|ccccc|}
   \hline
    $\Delta$ & $m$ & Tot. traces & \# samples & ratio\\
   \hline
      20 & 5 & 365818472 & 22774 & 16800\\
      10 & 5 & 106678064 & 20147 & 5295\\
      10 & 3 & 6196850 & 6552 & 946\\
       5 & 5 & 66947355 & 15937 & 4200\\
       3 & 5 & 23152990 & 15070 & 1536\\
  \hline
  \end{tabular}
\caption{The ratio between the total number of traces and
 the number of samples we would take using $C=10$.
 }\label{fig:time}
 \end{center}
\end{figure}

\begin{figure}
\begin{center}
\begin{tabular}{|c|c|c|}
	\hline
{$C$} & {\begin{tabular}{c}False negative\\ probability\end{tabular}} & {\begin{tabular}{c}Significantly false\\ positive probability\end{tabular}}\\
\hline
\hline
3 & 0.199 & 0.173\\
5 & 0.125 & 0.127\\
10 & 0.0671 & 0.0420\\
15 & 0.0180 & 0.0376\\
20 & 0.0108 & 0.0318\\
30 & 0.00195 & 0.0103\\
\hline
\end{tabular}
\caption{Probability that a trace with frequency $\varepsilon$ or more is not reported (false negative), and probability that a trace with frequency less than $\varepsilon/4$ is reported (significantly false positive), for different values of parameter $C$. The values are computed using the Poisson approximation to the binomial distribution, which is accurate unless the set $S_m$ from which we sample is small.}\label{table:errorprob}
%Source: http://stattrek.com/Tables/Poisson.aspx
\end{center}
\end{figure}

%\section{Open questions}
%\IEEEtriggeratref{6}
%There are a number of obvious open questions to ask in the context of this work.

\newpage

\bibliographystyle{plain}
\bibliography{spopos}

%\bibliography{spopos,../my}

\end{document}